\documentclass[man,12pt,floatsintext,natbib]{apa7} 
\usepackage{setspace,enumitem}   
\usepackage{amsmath,amsfonts,amsthm}
\usepackage{algorithm,algpseudocode}
\usepackage{multirow}
\usepackage{graphicx}
\usepackage{float} 
\usepackage{rotating}

\DeclareFontShape{T1}{lmr}{b}{it}{<->ssub*lmr/b/sl}{}
\DeclareRobustCommand{\bfseries}{\fontseries{b}\selectfont}
\DeclareMathAlphabet{\mathbf}{OT1}{lmr}{b}{n}

\definecolor{red1}{RGB}{228,55,55}
\definecolor{blue1}{RGB}{40,80,156}
\long\def\yl#1{\bgroup\color{blue1}#1\egroup}

\setlist{nosep,itemindent=*,leftmargin=*}

\def\t{^\top}
\def\dto{\stackrel{d}{\to}}
\def\pto{\stackrel{p}{\to}}

\def\ff{\mathbf{f}}
\def\gg{\mathbf{g}}
\def\hh{\mathbf{h}}
\def\ss{\mathbf{s}}

\def\yy{\mathbf{y}}
\def\ll{\boldsymbol{\ell}}

\def\EE{\mathbf{E}}
\def\HH{\mathbf{H}}
\def\UU{\mathbf{U}}
\def\YY{\mathbf{Y}}
\def\var{\mathrm{Var}}
\def\cov{\mathrm{Cov}}
\def\proj{\mathrm{proj}}
\def\real{\mathcal{R}}
\def\zero{\mathbf{0}}
\def\cF{\mathcal{F}}

\def\cN{\mathcal{N}}
\def\cQ{\mathcal{Q}}

\def\cU{\mathcal{U}}
\def\cY{\mathcal{Y}}
\def\sN{\mathsf{N}}
\def\sU{\mathsf{U}}
\def\pr{\mathbb{P}}
\def\ex{\mathbb{E}}
\def\bbeta{\boldsymbol{\beta}}

\def\ttheta{\boldsymbol{\theta}}

\def\eeta{\boldsymbol{\eta}}
\def\nnu{\boldsymbol{\nu}}
\def\pphi{\boldsymbol{\varphi}}
\def\ppsi{\boldsymbol{\psi}}
\def\PPhi{\boldsymbol{\Phi}}
\def\DDelta{\mathbf{\Delta}}

\def\OOmega{\mathbf{\Omega}}

\theoremstyle{plain}

\newtheorem{prop}{Proposition}
\theoremstyle{remark}
\newtheorem{rem}{Remark}
\makeatletter
\def\thm@space@setup{%
  \thm@preskip=0pt \thm@postskip=0pt
}
\def\cor@space@setup{%
  \thm@preskip=0pt \thm@postskip=0pt
}
\def\rem@space@setup{%
  \thm@preskip=0pt \thm@postskip=0pt
}
\def\proof@space@setup{%
  \thm@preskip=0pt \thm@postskip=0pt
}
\makeatother

\tikzset{
  mv/.style={thick,draw,rectangle,minimum width=18pt, minimum height=10pt},
  lv/.style={thick,draw,circle,minimum width=22pt, minimum height=22pt}
}

\title{\boldmath Two-stage Estimation of Latent Variable Regression Models: A General, Root-$N$ Consistent Solution}
\shorttitle{Two-Stage Estimation}
\authorsnames[1,2,2,1,2,2,3]{Yang Liu, Xiaohui Luo, Jieyuan Dong, Youjin Sung, Yueqin Hu, Hongyun Liu, Daniel J. Bauer}
\authorsaffiliations{{University of Maryland, College Park},{Beijing Normal University},{The University of North Carolina at Chapel Hill}}
\authornote{Correspondence should be addressed to Yang Liu (11112020106@bnu.edu.cn) at 3942 Campus Dr, College Park, MD 20742, USA, or Hongyun Liu (hyliu@bnu.edu.cn) at 19 Xinjiekouwai St. Beijing 100875, China.}
 
\abstract{Latent variable (LV) models are widely used in psychological research to investigate relationships among unobservable constructs. When one-stage estimation of the overall LV model is challenging, two-stage factor score regression (FSR) serves as a convenient alternative: the measurement model is fitted to obtain factor scores in the first stage, which are then used to fit the structural model in the subsequent stage. However, naive application of FSR is known to yield biased estimates of structural parameters. In this paper, we develop a generic bias-correction framework for two-stage estimation of parametric statistical models and tailor it specifically to FSR. Unlike existing bias-corrected FSR solutions, the proposed method applies to a broader class of LV models and does not require computing specific types of factor scores. We establish the $\sqrt{n}$-consistency of the proposed bias-corrected two-stage estimator under mild regularity conditions. To ensure broad applicability and minimize reliance on complex analytical derivations, we introduce a stochastic approximation algorithm for point estimation and a Monte Carlo-based procedure for variance estimation. In a sequence of Monte Carlo experiments, we demonstrate that the bias-corrected FSR estimator performs comparably to the ``gold standard'' one-stage maximum likelihood estimator. These results suggest that our approach offers a straightforward yet effective alternative for estimating LV models.}
\keywords{latent variable models, structural equation models, factor analysis, item response theory, factor score regression, asymptotic theory, stochastic approximation, numerical differentiation}

\begin{document}
\maketitle

\section{Introduction}
A prominent feature distinguishing psychological sciences from physical sciences is that many key variables of interest cannot be directly observed; these unobservable variables are typically referred to as \emph{constructs} \citep[e.g.,][Chapter 1]{CronbachMeehl1955, DeBoeckEtAl2023, EdwardsBagozzi2000, RaykovMarcoulides2011}. Latent variable (LV) models have become standard statistical operationalizations of psychological theories about constructs \citep[e.g.,][]{Bollen1989, Muthen2002, SkrondalRH2004}. Typically, a LV model can be decomposed into a \emph{measurement model} and a \emph{structural model} (\citealp{BartholomewEtAl2011}, Chapter 1; \citealp{Bollen1989}, Chapter 8). The measurement model characterizes the association between LVs, which are mathematical representations of constructs (e.g., emotional distress, intelligence, etc.), and manifest variables (MVs), which are observable indicators (e.g., responses to questionnaire or test items, response times, etc.); this model fully specifies the conditional distribution of MVs given LVs. Meanwhile, the structural model relates the LVs by a (potentially nonlinear) regression or path model. While the measurement model is essential for establishing the meaning of LVs, the ultimate goal for psychologists is often to draw inference about parameters in the structural model, which directly formulates the substantive theory about the target constructs.

The ``gold standard'' approach for estimating LV models is maximum likelihood \citep[ML;][]{Fisher1922}. As a \emph{one-stage} method, ML estimation entails a simultaneous search for both the measurement and structural parameters by maximizing the (log-)likelihood function. When the model is correctly specified and sufficiently regular, the ML estimator is known to be unbiased, normal, and efficient in large samples \citep[e.g.,][]{BickelDoksum2015}. In reality, however, one-stage estimation may not be desirable, or even feasible, for multiple reasons. First, one could encounter numerical glitches (e.g., non-convergence, ill-conditioned log-likelihood surfaces, etc.) or invalid solutions (e.g., Heywood cases) when performing one-stage ML in small to medium samples \citep[e.g.,][]{ChenEtAl2001, Rindskopf1984, vanDriel1978, WangEtAl2023}. Second, for complex LV models (e.g., multidimensional item response theory [IRT] models; \citealp{LiuEtAl2018, Reckase2009}), the likelihood function itself is often an intractable integral. Consequently, ML estimation must rely on sophisticated numerical integration or search techniques, which in turn demand careful tuning and can be computationally costly \citep[e.g.,][]{Cai2010a, Cai2010b, Haberman2006, SchillingBock2005, ZhangChen2022, ZhangEtAl2020}. Third, despite providing better representation of psychological theory, a fully customized LV model may exceed the capacity of existing software packages, such as M\emph{plus} \citep{mplus}, {\tt lavaan} \citep{lavaan}, \emph{mirt} \citep{mirt}, GLLAMM \citep{gllamm}, etc. Developing computer code for ML estimation from scratch is a task that is challenging, if not unrealistic, for most psychologists. Fourth, due to practical concerns such as test security and data portability, some public-use datasets only contain composite scores (e.g., scale scores) rather than raw MV-level data (e.g., responses to test items). Consequently, this precludes the use of one-stage ML estimation for the full LV model.

Two-stage estimation serves as a practical alternative when one-stage ML is not viable. It follows a ``divide and conquer'' principle by partitioning model parameters into two disjoint subsets: \emph{focal} parameters, which often include the key effects of interest, and \emph{nuisance} parameters, which collect all the remaining ones. These two subsets of parameters are estimated subsequently: we estimate the nuisance parameters first and then the focal parameters conditional on the first-stage nuisance parameter estimates. In the current paper, we focus mainly on a common two-stage method in the context of LV modeling: namely, \emph{factor score regression} \citep[FSR;][]{DevliegerEtAl2016, LastovickaThamodaran1991, SkrondalLaake2001}.\footnote{The statistical theory and computational techniques we develop, however, can be broadly extended to other two-stage estimation problems. See the ``Discussion'' section for more details.} The first stage of FSR involves estimating all the measurement parameters together with part of the structural parameters in order to predict factor scores. In the second stage of FSR, the remaining structural parameters of the LV model are estimated by fitting regressions to the predicted factor scores as if they were the LVs themselves.\footnote{Here, we do not mean to downplay the importance of measurement modeling by labeling measurement parameters as ``nuisance.'' The terms ``nuisance'' versus ``focal'' primarily reflect the order in which those parameters are estimated.} Additional two-stage estimation methods of LV models exist, such as the ``structural after measurement'' \citep{RosseelLoh2024, RosseelEtAl2025} and the pseudo ML approach \citep{KuhaBakk2025, LiuEtAl2019}. A brief review of these methods can be found in the subsection ``Existing Bias Correction Methods for Factor Score Regression.''

FSR offers two advantages: one computational and one substantive. On the computational end, FSR is often much more stable than one-stage ML, particularly with limited sample sizes \citep[e.g.,][]{DevliegerRosseel2022}. One reason is that the measurement model is often decomposable into simpler sub-models (e.g., one-dimensional factor models), which can be separately and efficiently estimated. Another reason is that the second-stage regression with factor scores is typically easy to solve; for all the examples in this paper, analytical solutions can be obtained without invoking any iterative algorithms (e.g., by ordinary least squares). On the substantive end, FSR is appealing to psychologists because it mirrors the established protocols for measure development. Psychologists typically calibrate and validate scales before using them to study associations among constructs \citep[e.g.,][]{AndersonGerbing1988, std2014}. Treating factor scores from validated measures as observed inputs allows FSR to streamline the analysis, allowing researchers to efficiently evaluate various structural models.

However, two-stage estimation suffers from a major drawback: the estimated focal parameters are generally biased, which can mislead statistical inference. A well-known example is the attenuation of a simple linear regression slope when the predictor variable is measured with error (e.g.,\citealp{Bollen1989}, pp. 154-159; \citealp{BollenSchwing1987}). When the structural model is more complex, substituting LVs by factor scores can result in either attenuation or amplification of estimated structural parameters, and the magnitude of bias may depend on the true parameter values (e.g., \citealp{Bollen1989}, pp. 159-178; see also \citealp{DeNadaiEtAl2022} for related discussions in psychiatric research). Although specific solutions exist for linear structural equation models \citep[SEMs; e.g.,][]{Croon2002, RosseelLoh2024, SkrondalLaake2001}, and latent class models \citep[e.g.,][]{BakkKuha2021, BakkEtAl2013, BolckEtAl2004}, general-purpose bias-correction methods for FSR are lacking. An additional challenge arises from the need to adjust the second-stage standard errors (SEs) of structural parameters, accounting for the sampling error carried over from the first-stage estimates of nuisance parameters \cite[e.g.][]{CanRosseel2025}.

To close this gap, we propose a general bias-correction strategy for two-stage estimation in models with multiple focal and nuisance parameters. Our method generalizes the work of \citet{LeungWang1998}, which is limited to bias correction in one-parameter models (i.e., with a single focal parameter and no nuisance parameter). We establish the $\sqrt{n}$-consistency of the bias-corrected estimator, where $n$ denotes the sample size. We also develop practical algorithms to obtain bias-corrected point estimators and valid large-sample SEs for structural parameters. A notable feature of our algorithm is the minimal reliance on analytics. It only requires users' input of two computer programs: one to compute the initial (potentially biased) two-stage estimator and the other to generate data based on the full LV model. We demonstrate the effectiveness of our proposed method by applying it to FSR across various common LV models.

The rest of the paper is structured as follows. We begin with a general statistical formulation of two-stage estimation, focusing on its application to FSR. We also briefly review existing $\sqrt{n}$-consistent methods for two-stage estimation of LV models. We then introduce our general bias-correction strategy, establish its $\sqrt{n}$-consistency, and provide practical algorithms for computing point estimates and standard errors (SEs). To evaluate the performance of the proposed estimator, we conduct three simulation studies concerning simple latent regression, latent moderation analysis, and multidimensional IRT, respectively. 
We conclude with a discussion of key takeaways, practical guidance, limitations, and directions for future research.

\section{Statistical Framework of Two-Stage Estimation}

\subsection{Two-Stage Estimation of Statistical Models}

Let $\YY$ be random data taking values in the data space $\cY$. We assume the data generating model $\pr_{\ttheta^*}$ is governed by the $q\times 1$ true parameter vector $\ttheta^*$, which belongs to a $q$-dimensional parameter space $\cQ\subseteq\real^q$. Additionally, we denote an observed realization of $\YY$ by $\yy\in\cY$ and a general parameter vector from $\cQ$ by $\ttheta$. We partition the parameter vector as $\ttheta = (\nnu\t, \pphi\t)\t$, which induces a corresponding partition of the parameter space $\cQ = \cN\times\cF$. Here, $\nnu \in \cN \subseteq \real^{q_0}$ ($q_0 < q$) are the \emph{nuisance parameters}, and $\pphi \in \cF \subseteq \real^{q_1}$ ($q_1 = q - q_0$) are the \emph{focal parameters}. In most problems, $\pphi$ is chosen to include the key effect sizes of interest, while $\nnu$ collects the remaining model parameters in $\ttheta$.
 
Given observed data $\yy$, a general two-stage estimation procedure involves: (1) obtaining the nuisance parameter estimates $\hat\nnu(\yy)$, and (2) obtaining the focal parameter estimates $\hat\pphi(\yy; \hat\nnu(\yy))$ given $\hat\nnu(\yy)$. Our notation for the second-stage estimator of the focal parameters, $\hat\pphi:\cY\times\cN\to\cF$, highlights its potential dependency on the nuisance parameters. Two-stage estimation is favorable provided both estimators, $\hat\nnu(\yy)$ and $\hat\pphi(\yy; \hat\nnu(\yy))$, can be efficiently computed. In practice, this efficiency is often achieved in two ways. First, the nuisance parameter estimator $\hat\nnu(\yy)$ may be estimated in a piecewise (and thus faster) manner. Second, we may reduce raw data to lower-dimensional summary statistics given $\hat\nnu(\yy)$, denoted $\ss(\yy; \hat\nnu(\yy))$, and subsequently obtain the focal parameter estimator $\hat\pphi(\yy; \hat\nnu(\yy))$ from only the reduced data $\ss(\yy; \hat\nnu(\yy))$. Both computational shortcuts are applicable to FSR, which we discuss next.

\subsection{Factor Score Regression}

In LV modeling, the data $\YY$ consist of independent and identically distributed (i.i.d.) MV vectors from $n$ individuals, denoted $\YY_1, \dots, \YY_n$. For each $i = 1,\dots, n$, $\YY_i = (Y_{i1}, \dots, Y_{ip})\t\in\real^p$. To highlight the role of the sample size $n$, we will write $\YY_{1:n}$ and $\yy_{1:n}$ in place of $\YY$ and $\yy$ in the sequel. In addition, let $\eeta_1, \dots, \eeta_n$ be i.i.d. LV vectors, where each $\eeta_i = (\eta_{i1}, \dots, \eta_{id})\t\in\real^d$ and $d < p$. The true model for the MV vector $\YY_i$, common across all $i = 1,\dots, n$, is characterized by two components. First, the \emph{structural model} defines the distribution of the LV vector $\eeta_i$. Second, the \emph{measurement model} determines the conditional distribution of the MV vector $\YY_i$ given the LV vector $\eeta_i$. The structural and measurement models together imply the joint distribution of $\YY_i$ and $\eeta_i$, which in turn induces the marginal distribution of the observable $\YY_i$.

For FSR, the nuisance parameter vector $\nnu$ comprises all the measurement parameters as well as those structural parameters needed for model identification (e.g., LV variances). Meanwhile, the focal parameter vector $\pphi$ collects the remaining structural parameters. Upon observing the data $\yy_{1:n}$, FSR proceeds in two stages: (1) obtaining the nuisance parameter estimates $\hat\nnu(\yy_{1:n})$ and predicting individual factor scores $\ss_i(\yy_i; \hat\nnu(\yy_{1:n}))$, $i = 1, \dots, n$, and (2) obtaining the focal parameter estimates $\hat\pphi(\yy_{1:n}; \hat\nnu(\yy_{1:n}))$ from a regression using only factor scores.

In the first stage of FSR, the nuisance parameters $\nnu$ can often be further partitioned and estimated in a piecewise fashion. In all the examples of the present paper, the measurement components have independent-cluster factor structures; therefore, the nuisance parameters can be estimated by fitting a one-factor model for one LV at a time. One factor models, with either continuous or discrete MVs, tend to be computationally faster and more stable than the original LV model \cite[e.g.,][]{AndersonGerbing1984}. Even for general measurement models lacking an independent-cluster structure, it may still be possible to identify hierarchical factor structures or locally independent sub-models, allowing for a piecewise estimation strategy that avoids estimating all the nuisance parameters simultaneously. 

Using the estimated nuisance parameters, we proceed to predict factor scores for each LV. Common factor score predictors include, but are not limited to, the Bartlett method \citep{Bartlett1937} and the regression method \citep{Thomson1936, Thurstone1935}. A simpler alternative to factor scores is the summed (or mean) score, calculated by summing (or averaging) all the MVs that load positively on the LV.\footnote{With a slight abuse of notation, regression using summed (or mean) scores is also referred to as FSR in the present work. In addition, summed (or mean) scores computed this way are also referred to as sub-scores when multiple LVs are involved.} Summed (or mean) scores are generally less reliable than factor scores \citep[in fact, the regression factor score attains the maximal reliability;][]{Raykov2004}, but may offer greater robustness to sampling and model specification errors for certain applications \citep{LiuPek2024}. See \citet{McNeishWolf2020}, \citet{WidamanRevelle2023}, \citet{McNeish2023}, and \citet{WidamanRevelle2024} for recent debates on using factor versus summed (or mean) scores in psychological research.

The second stage of FSR entails regression analysis using predicted factor scores $\ss_i(\yy; \hat\nnu(\yy_{1:n}))$, $i = 1,\dots, n$, conditional on the estimated measurement parameters $\hat\nnu(\yy_{1:n})$. In many applications, including all the simulations in this paper, the regression is linear, allowing for ordinary least squares (OLS) estimation which is non-iterative and thus computationally fast. However, FSR is not restricted to structural models that are additive regressions. When the regression function is nonlinear, we may still predict the outcome factor scores from the predictor factor scores using nonlinear least squares \citep{BatesWatts1988}. Although the estimation of nonlinear regressions typically requires iterative numerical search, it remains substantially less burdensome and more stable than one-stage ML estimation of the original LV model. Further extensions of FSR to other structural models (e.g., path analysis and generalized linear models) are straightforward \citep[e.g.,][]{AnderssonYW2021, DevliegerRosseel2017}, although they are not pursued further in this paper.

\subsection{Existing Bias-Corrected Two-Stage Estimators for Latent Variable Models}

While FSR is known to produce biased focal parameter estimates in general \citep[e.g.,][]{DevliegerEtAl2016}, the estimator can be asymptotically valid for certain families of LV models. This requires carefully choosing factor-score types. Alternatively, the second-stage estimation can use other summary statistics or the raw data instead of predicted factor scores, potentially yielding $\sqrt{n}$-consistent estimators in both stages. We briefly review these relevant approaches next.

In linear factor analysis, the Anderson-Rubin-McDonald (ARM) factor scores are known to preserve the correlations among LVs (\citealp{AndersonRubin1956}; \citealp{McDonald1981}; see also \citealp{Tucker1971}). As such, structural parameters identifiable from the LVs' correlation structure (e.g., standardized regression coefficients) can be consistently recovered using the ARM factor scores \citep{BogaertEtAl2026}. Likewise, \citet{SkrondalLaake2001} demonstrated that specific choices of factor scores in FSR enable consistent estimation of the structural regression. They proved consistency, in particular, for the FSR estimator of latent regression coefficients when the Bartlett factor scores are used for the $y$-side LVs (i.e., outcomes) and the regression factor scores are used for the $x$-side LVs (i.e., predictors). However, restricting the use of factor scores is an unsatisfactory solution for two reasons. First, researchers select factor scores based on their specific statistical properties: for example, Bartlett factor scores are unbiased with minimal variance \citep[][Corollary 2]{KrijnenEtAl1996}, regression factor scores minimize mean squared error \citep[][Proposition 1]{KrijnenEtAl1996}, and summed scores are easy to compute. Maintaining multiple sets of factor scores for different inferential purposes is impractical and potentially confusing. Second, when the LV model extends beyond linear SEM, a factor score choice that guarantees consistent recovery of the structural parameters may not exist. For example, \citet{LuThomas2008} showed that \citeauthor{SkrondalLaake2001} (\citeyear{SkrondalLaake2001})'s bias-avoiding strategy is, in principle, generalizable to IRT measurement models. However, unlike Bartlett factor scores in linear factor analysis, there is typically no IRT factor score that satisfies the required sufficient condition of unbiasedness.

In FSR, the summary statistics $\ss(\yy_{1:n}; \hat\nnu(\yy_{1:n}))$ are chosen to be the predicted factor scores in FSR. It is also possible to construct two-stage estimators of LV models using other summary statistics or even raw data. For linear SEMs, the fundamental cause of FSR's bias is that the covariance matrix of factor scores does not coincide with that of LVs. Tackling this issue directly, \citet{Croon2002} derived a correction formula that recovers the true LV covariances from the factor-score covariances, allowing the structural parameters to be consistently estimated using these corrected covariances. A generalization of Croon's (\citeyear{Croon2002}) formula is termed the local SAM approach in \citet{RosseelLoh2024}. In both Croon's correction and the local SAM, the estimated LV covariance matrix serves as input (i.e., summary statistics) for the second-stage estimation of the focal parameters. \citet{RosseelEtAl2025} recently extended local SAM to accommodate latent interactions and polynomial regressions by passing an approximate covariance matrix of all additive terms regarding the LVs to the second stage.

To apply local SAM, the structural model must be identifiable from a covariance matrix of latent quantities (e.g., LVs and their polynomial terms), which in turn must be consistently estimable from the data. In contrast, pseudo ML estimation \citep{GongSamaniego1981, Parke1986}, which is also referred to as the global SAM approach \cite{RosseelLoh2024} and is subsumed under the method of generalized estimating equations with nuisance parameters \citep{YuanJennrich2000}, offers a completely general two-stage solution. For instance, \citet{KuhaBakk2025} applied pseudo ML to estimate general LV models. Similar to FSR, this method estimates nuisance parameters first. However, their second stages differ: pseudo ML fits the full LV model while fixing all the nuisance parameters at their first-stage estimates. In a slightly different context (i.e, restricted recalibration) where the nuisance and focal parameters are estimated sequentially from independent samples, \citet{LiuEtAl2019} applied pseudo ML to obtain asymptotically correct standard errors and goodness-of-fit tests. Connecting these to our general recipe for two-stage estimation, pseudo ML passes the raw data, rather than any summary statistics thereof, as the input for the second stage. The drawback of pseudo ML is that its second stage is essentially as demanding as one-stage ML, providing little practical gain when the one-stage ML itself is computationally challenging.

\subsection{A General Bias-Correction Strategy}
The above survey of the extant literature reveals the absence of a general (i.e., not limited to linear measurement and structural models) and computationally efficient (i.e., faster than one-stage ML) bias-correction strategy for FSR. We contend such methodology is in pressing need given the computational convenience of FSR and its popularity in psychological research. Inspired by the original work of \citet{LeungWang1998}, which is limited to one-parameter statistical models, we propose a ``blackbox'' bias-correction method via stochastic approximation \citep[SA;][]{KushnerYin2003, RobbinsMonro1951}. This method can be broadly applied to two-stage estimation of parametric statistical models with multiple nuisance and focal parameters, subsuming FSR as a special case. In what follows, we first illustrate the intuition behind the bias-correction strategy using a one-parameter example (i.e., the scenario investigated by \citealp{LeungWang1998}). Subsequently, we formally establish a proposition for the $\sqrt{n}$-consistency of the bias-corrected estimator.

\begin{figure}[!t]
  \centering
  \includegraphics[width=\textwidth]{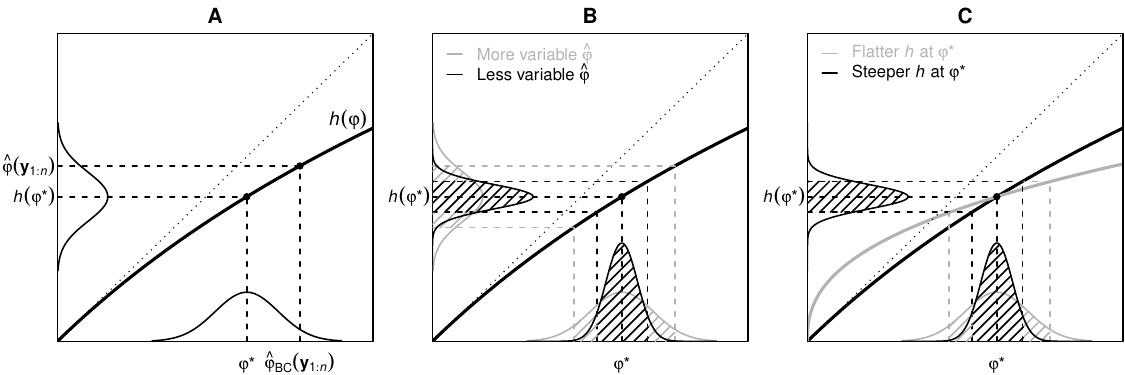}
  \caption{Graphical illustration of the bias-correction strategy in a single-parameter model. The parameter is denoted by $\varphi$. Panel A: The expected value of the initial (possibly biased) estimator $\hat\varphi(\YY_{1:n})$, $h(\varphi) = \ex_{\varphi}\hat\varphi(\YY_{1:n})$, is plotted as a monotonic function of $\varphi$ (shown as the thick curve). Given the observed data $\yy_{1:n}$, the bias-corrected estimate, denoted $\hat\varphi_{\rm BC}(\yy_{1:n})$, is obtained by mapping the initial estimate $\hat\varphi(\yy_{1:n})$ from the $y$-axis back to the $x$-axis using the curve $h$. The corresponding sampling distribution of the bias-corrected estimator (i.e., the bell-shaped curve on the $x$-axis) is induced by the sampling distribution of the initial estimator (i.e., the bell-shaped curve on the $y$-axis). Panel B: The bias-corrected estimator is more variable if the initial estimator is more variable, holding the rate of change of $h$ constant at the data generating $\varphi^*$. Panel C: The bias-corrected estimator is more variable if the function $h$ is flatter at $\varphi^*$, holding the variability of the initial estimator constant.}
  \label{fig:illus}
\end{figure}

Consider a model parameterized by a single (focal) parameter $\varphi$. Let $\hat\varphi(\YY_{1:n})$ be an initial estimator of $\varphi$, and $h(\varphi) = \ex_{\varphi}\hat\varphi(\YY_{1:n})$ the expectation of the initial estimator when the data generating parameter equals to $\varphi$. The initial estimator is then biased (at $\varphi$) when $h(\varphi)\ne \varphi$. \citeauthor{LeungWang1998}'s (\citeyear{LeungWang1998}) bias-correction strategy is applicable whenever $h(\varphi)$ is a one-to-one function of $\varphi$.\footnote{To establish asymptotic theory, it suffices to assume $h$ is one-to-one function in some neighborhood of the true $\varphi^*$.} A graphical illustration can be found in Figure \ref{fig:illus}A. Geometrically, knowing $h(\varphi)$ (i.e., the $y$-coordinate) is equivalent to knowing $\varphi$ (i.e., the $x$-coordinate) because the latter can be recovered from the graph of $h$: mathematically, $\varphi = h^{-1}(h(\varphi))$ where $h^{-1}$ is the unique inverse function of $h$. Similarly, the initial estimator, denoted $\hat\varphi(\YY_{1:n})$, can be mapped from the $y$-axis back to the $x$-axis using the same inverse function $h^{-1}$, yielding the \emph{bias-corrected estimator} $\hat\varphi_{\rm BC}(\YY_{1:n}) = h^{-1}(\hat\varphi(\YY_{1:n}))$. When the data generating parameter is $\varphi^*$, the initial estimator $\hat\varphi(\YY_{1:n})$, albeit biased for $\varphi^*$, is guaranteed to be unbiased and is typically consistent for $h(\varphi^*)$. Hence, the bias-corrected estimator $\hat\varphi_{\rm BC}(\YY_{1:n})$ is expected to concentrate around $\varphi^*$, particularly in large samples.

This graphical illustration also suggests that the efficiency of the bias-corrected estimator $\hat\varphi_{\rm BC}(\YY_{1:n})$ is governed by two factors: the variability of the initial estimator $\hat\varphi(\YY_{1:n})$ around $h(\varphi^*)$ and the instantaneous rate of change of the function $h$ at $\varphi^*$. On the one hand, a less precise initial estimator remains less precise after bias correction, holding the function $h$ constant (see Figure \ref{fig:illus}B). On the other hand, a flatter $h$ (i.e., one with a smaller gradient) at the true parameter $\varphi^*$ inflates the variance of the bias-corrected estimator, holding the variance of the initial estimator constant (see Figure \ref{fig:illus}C). To enhance the practical efficiency of bias-corrected estimators, one must select initial estimators judiciously to balance the two contributing factors. 

The aforementioned bias-correction heuristic generalizes to the estimation of $q_1$ focal parameters $\pphi$ in the presence of $q_0$ nuisance parameters $\nnu$. To adapt the notations to this more general scenario, let $\hh(\pphi; \nnu) = \ex_{\nnu, \pphi}\hat\pphi(\YY_{1:n}; \nnu)$ be the expected value of the initial focal parameter estimator $\hat\pphi(\YY_{1:n}; \nnu)$ under data generating parameter values $\ttheta = (\nnu\t, \pphi\t)\t$. Denote by $\hat\nnu(\YY_{1:n})$ the nuisance parameter estimator based on the same observed data as usual. We formally establish the $\sqrt{n}$-consistency of the bias-corrected estimator in the following proposition. While the proof is a straightforward application of the Delta method \citep[e.g.,][Lemma 5.3.3]{BickelDoksum2015}, implicit differentiation is required to derive the asymptotic covariance matrix (ACM).

\begin{prop}
  Suppose that the nuisance parameter estimator $\hat\nnu(\YY_{1:n})$ and the initial focal parameter estimator $\hat\pphi(\YY_{1:n}; \hat\nnu(\YY_{1:n}))$ satisfy 
  \begin{equation}
    \sqrt{n}\begin{bmatrix}
      \hat\nnu(\YY_{1:n}) - \nnu^*\\
      \hat\pphi(\YY_{1:n}; \hat\nnu(\YY_{1:n})) - \hh(\pphi^*; \nnu^*)
    \end{bmatrix}
    \dto\sN\left(\zero_q, \OOmega(\ttheta^*)\right),
    \label{eq:rootn0}
  \end{equation}
  under true model with parameters $\ttheta^* = (\nnu^*{}\t, \pphi^*{}\t)\t$,
  in which $\zero_q$ is a $q\times 1$ vector of zeros, and $\OOmega(\ttheta)$ is the $q\times q$ positive definite ACM for $\hat\ttheta(\YY_{1:n}) = (\hat\nnu(\YY_{1:n})\t, \hat\pphi(\YY_{1:n}; \hat\nnu(\YY_{1:n}))\t)\t$. Assume that the function $\hh$ is continuously differentiable with respect to both $\pphi$ and $\nnu$ in some neighborhood of $\ttheta^*$, in which the $q_1\times q_1$ Jacobian matrix $\nabla_{\pphi}\hh(\pphi; \nnu)$ is invertible. Let
  \begin{equation}
    \DDelta(\ttheta) = \left[-\nabla_{\pphi}\hh(\pphi; \nnu)^{-1}\nabla_{\nnu}\hh(\pphi; \nnu) : \nabla_{\pphi}\hh(\pphi; \nnu)^{-1}  \right].
    \label{eq:delta}
  \end{equation}
  Then the bias-corrected focal parameter estimator,
\begin{equation}
  \hat\pphi_{\rm BC}(\YY_{1:n}; \hat\nnu(\YY_{1:n})) = \hh^{-1}\left( \hat\pphi(\YY_{1:n}; \hat\nnu(\YY_{1:n})); \hat\nnu(\YY_{1:n}) \right),
  \label{eq:bcestim}
\end{equation}
is $\sqrt{n}$-consistent for $\pphi^*$ under the true model:
  \begin{equation}
    \sqrt{n}\left[ \hat\pphi_{\rm BC}(\YY_{1:n}; \hat\nnu(\YY_{1:n})) - \pphi^* \right]
    \dto \sN\left( \zero_{q_1}, \DDelta(\ttheta^*)\OOmega(\ttheta^*)\DDelta(\ttheta^*)\t\right).
    \label{eq:rootn}
  \end{equation}
  \label{prop:bc}
\end{prop}
\vspace*{-\baselineskip}
\begin{proof}
  Define the map $\ll(\pphi, \nnu, \ppsi) = \hh(\pphi; \nnu) - \ppsi$, where $\ppsi\in\real^{q_1}$. Clearly, $\ll(\pphi^*, \nnu^*, \hh(\pphi^*; \nnu^*)) = \zero_{q_1}$. By the continuous differentiability of $\hh$ (locally around $\ttheta^*$) and the Implicit Function Theorem \citep[e.g.,][Theorem 9.28]{Rudin1964}, the solution of $\pphi$ to the equation $\ll(\pphi, \nnu, \ppsi) = \zero_{q_1}$ can be uniquely represented by an implicitly defined function $\ff(\nnu, \ppsi)$ for $(\nnu\t, \ppsi\t)\t$ in some neighborhood of $(\nnu^*{}\t, \hh(\pphi^*; \nnu^*)\t)\t$. Moreover, the Jacobian matrices of $\hh$ are defined in the same neighborhood and can be expressed by differentiating both sides of the equation $\zero_{q_1} = \ll(\ff(\nnu, \ppsi), \nnu, \ppsi) = \hh(\ff(\nnu, \ppsi); \nnu) - \ppsi$ and rearranging terms:
  \begin{equation}
    \begin{aligned}
      \nabla_{\nnu}\ff(\nnu, \ppsi) &=  -\nabla_{\pphi}\hh(\ff(\nnu, \ppsi); \nnu)^{-1}\nabla_{\nnu}\hh(\ff(\nnu, \ppsi); \nnu),\\
      \nabla_{\ppsi}\ff(\nnu, \ppsi) &=  \nabla_{\pphi}\hh(\ff(\nnu, \ppsi); \nnu)^{-1}.\\
    \end{aligned}
    \label{eq:jac}
  \end{equation}
  When evaluating Equation \ref{eq:jac} at $\nnu = \nnu^*$ and $\ppsi = \hh(\pphi^*; \nnu^*)$, we have $\ff(\nnu^*, \hh(\pphi^*;\nnu^*)) = \pphi^*$ and thus $[\nabla_{\nnu}\ff(\nnu^*, \hh(\pphi^*; \nnu^*)) : \nabla_{\ppsi}\ff(\nnu^*, \hh(\pphi^*; \nnu^*))] = \DDelta(\ttheta^*)$ (i.e., Equation \ref{eq:delta}). Now, Equation \ref{eq:rootn0} implies $\hat\ttheta(\YY_{1:n})\pto(\nnu^*{}\t, \hh(\pphi^*; \nnu^*)\t)\t$ as $n\to\infty$. Therefore, 
  \begin{equation}
    \begin{aligned}
    &\sqrt{n}\left[ \hat\pphi_{\rm BC}(\YY_{1:n}; \hat\nnu(\YY_{1:n})) - \pphi^* \right]\\
    =\ &  \sqrt{n}\left[ \ff(\hat\nnu(\YY_{1:n}), \hat\pphi(\YY_{1:n}; \hat\nnu(\YY_{1:n}))) - \ff(\nnu^*, \hh(\pphi^*; \nnu^*)) \right] + o_p(1)\\
    \dto\ & \sN\left( \zero_{q_1}, \DDelta(\ttheta^*)\OOmega(\ttheta^*)\DDelta(\ttheta^*)\t\right)
    \end{aligned}
    \label{eq:asympequiv}
  \end{equation}
  by the Delta method. The proof is now complete.
\end{proof}

\begin{rem}
  When there is no nuisance parameter (i.e., $\pphi = \ttheta$ and $q_1 = q$), the notations for the initial estimator and its expectation can be simplified to $\hat\pphi(\YY_{1:n})$ and $\hh(\pphi) = \ex_{\pphi}\hat\pphi(\YY_{1:n})$, respectively (cf. the illustrative example in Figure \ref{fig:illus}). In this special case, Equation \ref{eq:delta} reduces to $\DDelta(\pphi) = \nabla_{\pphi}\hh(\pphi)^{-1}$. Subsequently, the ACM for the bias corrected estimator (see the right-hand side of Equation \ref{eq:rootn}) equals to $\nabla_{\pphi}\hh(\pphi^*)^{-1}\OOmega(\pphi^*)\nabla_{\pphi}\hh(\pphi^*)^{-\top}$, in which $\OOmega(\pphi^*)$ denotes the ACM for the initial estimator under the true model with parameters $\pphi^* = \ttheta^*$, and the superscript $^{-\top}$ is a shorthand for the transposed inverse. Examples of this special case include one-parameter models \citep{LeungWang1998} and differential equation models recently investigated by \citet{LuoEtAl2025}.
  \label{rmk:nonuis}
\end{rem}

\begin{rem}
  Equation \ref{eq:rootn0} requires that the initial estimators $\hat\nnu(\YY_{1:n})$  and $\hat\pphi(\YY_{1:n}; \hat\nnu(\YY_{1:n}))$ to be jointly $\sqrt{n}$-consistent for $\nnu^*$ and $\hh(\pphi^*; \nnu^*)$. This assumption is generally mild, given that these estimators are often derived by the method of estimating equations (e.g., ML and OLS; \citealp{Godambe1991}). For a thorough discussion of regularity conditions, please refer to \citet{YuanJennrich2000}.
  \label{rmk:init}
\end{rem}

\section{Algorithms}
This section details practical algorithms for approximate computation of the bias-corrected estimator (Equation \ref{eq:bcestim}) and its ACM (Equation \ref{eq:rootn}). To facilitate implementation, we present the pseudo-code for calculating point estimates and ACM in Algorithms \ref{alg:bc} and \ref{alg:acm}, respectively.

\subsection{Point Estimation}

\begin{algorithm}[!t]
  \caption{A Robbins-Monro Algorithm for Bias Correction}
  \begin{algorithmic}[1]
    \Require Starting values for focal parameters $\pphi^{(0)}$, nuisance parameter values $\nnu$, initial estimates for focal parameters $\hat\pphi(\yy_{1:n}; \nnu)$, number of iterations $K$, learning rate constants $a > 0$ and $b\in(0.5, 1]$
    \For{$k = 1, \dots, K$}
    \State Compute learning rate $\gamma_k = a k^{-b}$
    \State Generate random components $\UU^{(k)}$\label{ln:ugen}
    \State Generate data $\YY_{1:n}^{(k)} = \gg(\UU^{(k)}, (\hat\nnu\t, [\pphi^{(k - 1)}]\t)\t)$\label{ln:datagen}
    \State Update $\pphi^{(k)} = \pphi^{(k - 1)} - \gamma_k\left[ \hat\pphi(\YY^{(k)}_{1:n}; \nnu)- \hat\pphi(\yy_{1:n}; \nnu) \right]$\label{ln:rm}
  \EndFor
  \State Return $\pphi^{(1)}, \dots, \pphi^{(K)}$
  \end{algorithmic}
  \label{alg:bc}
\end{algorithm}

Given the observed data $\yy_{1:n}$, the bias-corrected focal parameter estimates (Equation \ref{eq:bcestim}) are obtained by finding the $\pphi$ values that solve the equation:
\begin{equation}
  \hh(\pphi; \nnu) = \ex_{\nnu, \pphi}\hat\pphi(\YY_{1:n}; \nnu) = \hat\pphi(\yy_{1:n}; \nnu),
  \label{eq:bceq}
\end{equation}
in which the nuisance parameters $\nnu$ are fixed at their sample estimates $\hat\nnu(\yy_{1:n})$. Equation \ref{eq:bceq} involves the intractable expectation $\hh$; however, it can be efficiently solved by standard SA provided that a noisy and unbiased estimator of $\hh$ is available. Let $\gg: \cU\times\cQ\to\cY$ be the \emph{data generating algorithm} (DGA), which formulates the procedure for simulating data from the population model. More specifically, $\gg$ maps the random components $\UU\in\cU$ with a completely known distribution and model parameters $\ttheta\in\cQ$ to sample data $\YY_{1:n} = \gg(\UU, \ttheta)\sim\pr_{\nnu, \pphi}$. It follows that $\hat\pphi(\gg(\UU, \ttheta); \nnu)$ is an unbiased estimator of $\hh(\pphi; \nnu)$. A standard Robbins-Monro (\citeyear{RobbinsMonro1951}) algorithm using this estimator is summarized as Algorithm \ref{alg:bc}. We make several technical remarks regarding the algorithm below.

\begin{rem}
  Algorithm \ref{alg:bc} returns the entire history of Robbins-Monro iterates after running a fixed budget of $K$ iterations, which facilitates convergence diagnostics. Various methods exist to derive final bias-corrected estimates for the focal parameters. The classic approach relies on the convergence of the Robbins-Monro iterates to the solution of Equation \ref{eq:bceq} as $K\to\infty$, which is ensured by the choice of $b\in(0.5, 1]$. By choosing a sufficiently large $K$, the final iterate $\pphi^{(K)}$ can be made arbitrarily close to the solution with high probability. Throughout the rest of this paper, however, we approximate the bias-corrected estimates by the learning-rate weighted average:
  \begin{equation}
    \bar\pphi^{(K)} = \frac{\sum_{k=\lfloor K/2\rfloor + 1}^K\gamma_k\pphi^{(k)}}{\sum_{k=\lfloor K/2\rfloor + 1}^K\gamma_k}.
    \label{eq:avg}
  \end{equation}
  Equation \ref{eq:avg} is computed after discarding the first half of iterations to mitigate the influence of the starting state $\pphi^{(0)}$. The weighted average $\bar\pphi^{(K)}$ is known to exhibit greater robustness and faster convergence than the final iterate $\pphi^{(K)}$ when the learning rate decays slowly \citep[e.g.,][]{NemirovskiEtAl2009}.
  \label{rmk:cvg}
\end{rem}

\begin{rem}
  In practice, we recommend setting $b\approx .5$ to ensure a slow decay of the learning rate, thereby avoiding spurious convergence. For instance,  we adopted $b = .6$ for all the numerical studies in this paper. The choice of the multiplier $a > 0$ is likely problem-specific. An excessively large $a$ value may cause numerical instability during the initial iterations, whereas a small $a$ may result in slow convergence.
  \label{rmk:tune}
\end{rem}

\begin{rem}
  When the parameter space $\cQ$ is bounded, it is often recommended to modify Line \ref{ln:rm} in Algorithm \ref{alg:bc} to
  \begin{equation}
    \pphi^{(k)} = \proj_{\cQ}\left(\pphi^{(k - 1)} - \gamma_k\left[ \hat\pphi(\YY^{(k)}_{1:n}; \nnu)- \hat\pphi(\yy_{1:n}; \nnu) \right]\right),
    \label{eq:rm}
  \end{equation}
  in which $\proj_{\cQ}$ stands for orthogonal projection onto $\cQ$. Equation \ref{eq:rm} safeguards the algorithm against invalid iterates. Provided the solution resides in the interior of $\cQ$, the resulting projected Robbins-Monro algorithm remains convergent under standard regularity conditions \citep[e.g., see Sections 5.1 and 5.2 of][]{KushnerYin2003}. In scenarios where $\proj_{\cQ}$ cannot be computed exactly (e.g., the third simulation study), other computationally efficient retraction operations onto the feasible set can be substituted and typically perform well in practice (see the third simulation study in the present work).
  \label{rmk:proj}
\end{rem}

\begin{rem}
  Although $\hat\pphi(\YY_{1:n}; \nnu)$, $\YY_{1:n}\sim\pr_{\nnu, \pphi}$, is an unbiased estimator of $\hh(\pphi; \nnu)$, its variance can be large when the sample size $n$ is limited. A simple variance reduction strategy entails simulating more than one set of data and estimating $\hh(\pphi; \nnu)$ by the Monte Carlo average of initial estimates. That said, our pilot simulations indicate that simulating a single set of data in each replication (Lines \ref{ln:ugen} and \ref{ln:datagen} in Algorithm \ref{alg:bc}) suffices for stable convergence across all the modeling contexts considered in this paper.
  \label{rmk:mult}
\end{rem}

\begin{rem}
  Lines \ref{ln:ugen} and \ref{ln:datagen} in Algorithm \ref{alg:bc} can be alternatively stated as a single step: simulate $\yy_{1:n}^{(k)}$ from the statistical model $\pr_{\hat\nnu(\yy_{1:n}), \pphi^{(k - 1)}}$. Although explicitly identifying the random components $\UU$ (using a DGA) is not strictly required for obtaining bias-corrected point estimates (Algorithm \ref{alg:bc}), it becomes critical for variance reduction when approximating the ACM of the biased-corrected estimator (Algorithm \ref{alg:acm}). We also note that the same statistical model can often be represented by more than one DGAs. For LV models in this work, we choose algorithms in which the random components $\UU$ are composed of independent $\sN(0, 1)$ and/or $\sU(0, 1)$ variates to map directly to LVs and regression error. Concrete examples can be found in the ``Simulation Studies'' section.
  \label{rmk:dga}
\end{rem}

\subsection{Asymptotic Covariance Matrix}

\begin{algorithm}[!t]
  \caption{Asymptotic Covariance Matrix by Simultaneously-Perturbation Monte Carlo}
  \begin{algorithmic}[1]
    \Require Parameter values $\ttheta = (\nnu\t, \pphi\t)\t$, number of replications $M$, perturbation constant $\delta > 0$
    \For{$m = 1, \dots, M$}
      \State Generate random components $\UU^{(m)}$
      \State Obtain initial estimate $\hat\ttheta(\gg(\UU^{(m)}, \ttheta))$
      \State Generate $\EE^{(m)}$ and evaluate $\HH_{\delta}^{(m)}(\ttheta)$ (Equation \ref{eq:spgrad})
  \EndFor
  \State Compute parametric bootstrap ACM $\hat\OOmega(\ttheta)$ from the initial estimates
  \State Compute SP Jacobian $\widehat{\nabla_{\ttheta}}\hh_\delta(\pphi; \nnu)$
  \State Compute $\hat\DDelta(\ttheta)$ from the SP Jacobian by Equation \ref{eq:delta}
  \State Return $\hat\DDelta(\ttheta)\hat\OOmega(\ttheta)\hat\DDelta(\ttheta)\t$
  \end{algorithmic}
  \label{alg:acm}
\end{algorithm}

By Proposition \ref{prop:bc}, the ACM for the bias-corrected estimator is given by $\DDelta(\ttheta^*)\OOmega(\ttheta^*)\DDelta(\ttheta^*)\t$. Subsequently, SEs are obtained by scaling the ACM by $n$ and taking the square roots of the diagonal elements. To approximate the ACM given sample data $\yy_{1:n}$, we replace the true $\ttheta^* = (\nnu^*{}\t, \pphi^*{}\t)\t$ by the consistent, bias-corrected estimates $\hat\ttheta_{\rm BC}(\yy_{1:n}) = (\hat\nnu(\yy_{1:n})\t, \hat\pphi_{\rm BC}(\yy_{1:n}; \hat\nnu(\yy_{1:n}))\t)\t$. We then seek Monte Carlo estimates for $\OOmega(\hat\ttheta_{\rm BC}(\yy_{1:n}))$ (i.e., the initial estimator's ACM) and $\DDelta(\hat\ttheta_{\rm BC}(\yy_{1:n}))$ (i.e., the Jacobian). It is worth noting that closed-form expressions may exist for these matrices, particularly for $\OOmega$; however, Monte Carlo estimation is favored here for its adaptability and easy implementation across a wide range of problems.

Algorithm \ref{alg:acm} computes both component matrices within a single loop of $M$ replications. Within each replication $m$, all data sets are generated using a shared collection of random components $\UU^{(m)}$ (via the DGA), an effective variance reduction technique known as the method of common random numbers \citep[e.g.,][]{Glynn1985, HeidelbergerIglehart1979}. For the initial estimator's ACM $\OOmega(\ttheta)$, Algorithm \ref{alg:acm} yields the usual parametric bootstrap estimate \citep[][Chapter 6]{EfronTibshirani1993}, denoted $\hat\OOmega(\ttheta)$. This corresponds to to the Monte Carlo sample covariance matrix computed from the replicates $\hat\ttheta(\gg(\UU^{(m)}, \ttheta))$, $m = 1, \dots, M$. Computing the matrix $\DDelta(\ttheta)$, which depends on the Jacobian matrices $\nabla_{\nnu}\hh(\pphi; \nnu)$ and $\nabla_{\pphi}\hh(\pphi; \nnu)$, poses a greater challenge. To address this, Algorithm \ref{alg:acm} employs the method of simultaneous perturbation \citep[SP;][]{Spall1992}, a Monte Carlo finite difference method for numerical differentiation. Formally, the SP estimator for $\nabla_{\ttheta}\hh(\pphi; \nnu)$ can be expressed by $\widehat{\nabla_{\ttheta}}\hh_\delta(\pphi; \nnu) = M^{-1}\sum_{m = 1}^M\HH_\delta^{(m)}(\ttheta)$ where
\begin{equation}
  \begin{aligned}
    \HH^{(m)}_\delta(\ttheta) =&\ (2\delta)^{-1}\Big[\hat\pphi(\gg(\UU^{(m)}, \ttheta + \delta\EE^{(m)}); \nnu + \delta\EE_{\nnu}^{(m)})\\
    &\ - \hat\pphi(\gg(\UU^{(m)}, \ttheta - \delta\EE^{(m)}); \nnu - \delta\EE_{\nnu}^{(m)})\Big] \left[\EE^{(m)}\right]^{-\top},\ m = 1, \dots, M.
  \end{aligned}
  \label{eq:spgrad}
\end{equation}
In Equation \ref{eq:spgrad}, $\delta > 0$ is the perturbation constant, and $\EE^{(m)} = (\EE_{\nnu}^{(m)}{}\t, \EE_{\pphi}^{(m)}{}\t)\t$, where $\EE_{\nnu}^{(m)}$ and $\EE_{\pphi}^{(m)}$ denote $q_0\times 1$ and $q_1\times 1$ independent Rademacher random vectors (i.e., each variate takes values $-1$ or 1 with 50/50 chance). Additionally, we denote by $[\EE^{(m)}]^{-1}$ the $q\times 1$ elementwise reciprocal of $\EE^{(m)}$, and by $[\EE^{(m)}]^{-\top}$ the transpose thereof. By Lemma 1 of \citet{Spall1992}, the bias and variance of $\widehat{\nabla_{\ttheta}}\hh_\delta(\pphi; \nnu)$ are of order $O(\delta^2)$ and $O(M^{-1})$, respectively. To secure an accurate and precise approximation of the ACM, we typically use a small perturbation constant (e.g., $\delta = 10^{-4}$) and a large number of replications (e.g., $M = 1000$) in practice.

\begin{rem}
  A key advantage of SP Jacobian estimation is that the number of evaluations of the initial estimator $\hat\pphi$ does not grow with the dimension of the parameter space $q$. However, pilot simulations revealed that for complex models (e.g., in the third simulation study), a large $M$ is needed to obtain a precise estimate of the ACM. In such scenarios, we recommend partitioning $\ttheta$ into $B$ blocks, denoted $\ttheta = (\ttheta_1\t, \dots, \ttheta_B\t)\t$ and perform SP in a blockwise fashion. \footnote{This partition is unrelated to the partition into nuisance and focal parameters and is introduced solely for the purpose of Jacobian approximation.} Empirically, we found that dividing $\ttheta$ into 10 to 20 equal-sized blocks often strikes a good balance between estimation precision and computation speed. Notably, if $B = q$ where each block contains a single parameter, the resulting SP estimate coincides with the central difference estimate applied to the Monte Carlo average $M^{-1}\sum_{m = 1}^M\hat\pphi(\gg(\UU^{(m)}, \ttheta); \nnu)$.
  \label{rmk:blk}
\end{rem}

\section{Simulation Studies}

In this section, we evaluate the performance of the bias-corrected FSR estimator and its corresponding SEs in a sequence of three simulation studies of increasing  model complexity. Our primary focus is to assess the empirical bias and variability of the proposed estimator in comparison with the naive FSR (without bias correction) and the one-stage ML. To this end, we report relative bias (RB) and empirical SE (ESE) for each candidate point estimator, as well as the relative bias in SE (RBSE) for all the $\sqrt{n}$-consistent estimators in theory.\footnote{Relative bias in SE is computed as the difference between the root mean asymptotic variance and the empirical SE, scaled by the empirical SE.} All reported summary statistics were aggregated across 500 replications. Unless otherwise specified, all computations were implemented in R version 4.5.1 \citep{CRAN}.

\begin{figure}[!t]
  \centering
  \begin{tikzpicture}[baseline,>=latex, scale=0.9]
    \node at (1, 2.5){\textsf{\bfseries A}};
    \node[lv] (x) at (1, 0){};
    \node[lv] (z) at (1, -3){};
    \draw[thick,->] (x) -- (z);
    \node[mv,black!30] (x1) at (0, 1.5){};
    \node[draw=none,black!30] (x2) at (1, 1.5){\ldots};
    \node[mv,black!30] (x3) at (2, 1.5){};
    \draw[thick,->,black!30] (x) -- (x1.south);
    \draw[thick,->,black!30] (x) -- (x3.south);
    \node[mv,black!30] (z1) at (0, -4.5){};
    \node[draw=none,black!30] (z2) at (1, -4.5){\ldots};
    \node[mv,black!30] (z3) at (2, -4.5){};
    \draw[thick,->,black!30] (z) -- (z1.north);
    \draw[thick,->,black!30] (z) -- (z3.north);
    \draw[thick,->] (0.0, -3) -- (z);
    \draw[thick,<->,black!30] (x) to [out=-20,in=20,loop,distance=15pt] (x);
  \end{tikzpicture}
  \hspace*{0.7in}
  \begin{tikzpicture}[baseline,>=latex, scale=0.9]
    \node at (1, 2.5){\textsf{\bfseries B}};
    \node[lv] (x) at (1, 0){};
    \node[lv] (w) at (2, -1.5){};
    \node[lv] (z) at (1, -3){};
    \draw[thick,-latex] (x) -- (z);
    \draw[thick,-latex] (w) -- (1, -1.5);
    \node[mv,black!30] (x1) at (0, 1.5){};
    \node[draw=none,black!30] (x2) at (1, 1.5){\ldots};
    \node[mv,black!30] (x3) at (2, 1.5){};
    \draw[thick,-latex,black!30] (x) -- (x1.south);
    \draw[thick,-latex,black!30] (x) -- (x3.south);
    \node[mv,black!30] (z1) at (0, -4.5){};
    \node[draw=none,black!30] (z2) at (1, -4.5){\ldots};
    \node[mv,black!30] (z3) at (2, -4.5){};
    \draw[thick,-latex,black!30] (z) -- (z1.north);
    \draw[thick,-latex,black!30] (z) -- (z3.north);
    \draw[thick,-latex] (0.0, -3) -- (z);
    \node[mv,black!30] (w1) at (3, -0.8){};
    \node[draw=none,black!30] (w2) at (3, -1.4){$\vdots$};
    \node[mv,black!30] (w3) at (3, -2.2){};
    \draw[thick,-latex,black!30] (w) -- (w1.west);
    \draw[thick,-latex,black!30] (w) -- (w3.west);
    \draw[thick,latex-latex] (x) ..controls (1.8, -0.2) and (2, -0.4).. (w);
    \draw[thick,latex-latex,black!30] (x) to [out=160,in=200,loop,distance=15pt] (x);
    \draw[thick,latex-latex,black!30] (w) to [out=250,in=290,loop,distance=15pt] (w);
  \end{tikzpicture}
  \hspace*{0.3in}
  \begin{tikzpicture}[baseline,>=latex, scale=0.9]
    \node at (1, 2.5){\textsf{\bfseries C}};
    \node[lv] (xa) at (1, 0.9375){};
    \node[lv] (xb) at (1, -0.6875){};
    \node[lv] (xc) at (1, -2.3125){};
    \node[lv] (xd) at (1, -3.9375){};
    \node[lv] (z) at (2, -1.5){};
    \draw[thick,-latex] (xa) -- (z);
    \draw[thick,-latex] (xb) -- (z);
    \draw[thick,-latex] (xc) -- (z);
    \draw[thick,-latex] (xd) -- (z);
    \node[mv,black!30] (x1) at (-0.5, 1.5){};
    \node[draw=none,black!30] (x2) at (-0.5, 1){$\vdots$};
    \node[mv,black!30] (x3) at (-0.5, 0.375){};
    \node[mv,black!30] (x4) at (-0.5, -0.125){};
    \node[draw=none,black!30] (x5) at (-0.5, -0.625){$\vdots$};
    \node[mv,black!30] (x6) at (-0.5, -1.25){};
    \node[mv,black!30] (x7) at (-0.5, -1.75){};
    \node[draw=none,black!30] (x8) at (-0.5, -2.25){$\vdots$};
    \node[mv,black!30] (x9) at (-0.5, -2.875){};
    \node[mv,black!30] (x10) at (-0.5, -3.375){};
    \node[draw=none,black!30] (x11) at (-0.5, -3.875){$\vdots$};
    \node[mv,black!30] (x12) at (-0.5, -4.5){};
    \draw[thick,-latex,black!30] (xa) -- (x1.east);
    \draw[thick,-latex,black!30] (xa) -- (x3.east);
    \draw[thick,-latex,black!30] (xb) -- (x4.east);
    \draw[thick,-latex,black!30] (xb) -- (x6.east);
    \draw[thick,-latex,black!30] (xc) -- (x7.east);
    \draw[thick,-latex,black!30] (xc) -- (x9.east);
    \draw[thick,-latex,black!30] (xd) -- (x10.east);
    \draw[thick,-latex,black!30] (xd) -- (x12.east);
    \node[mv,black!30] (z1) at (3, -0.8){};
    \node[draw=none,black!30] (z2) at (3, -1.4){$\vdots$};
    \node[mv,black!30] (z3) at (3, -2.2){};
    \draw[thick,-latex,black!30] (z) -- (z1.west);
    \draw[thick,-latex,black!30] (z) -- (z3.west);
    \draw[thick,latex-latex] (xa.west) -- (0.2, 0.9375) -- (0.2, -3.9375) -- (xd.west);
    \draw[thick,-latex] (0.2, -0.6875) -- (xb.west);
    \draw[thick,-latex] (0.2, -2.3125) -- (xc.west);
    \draw[thick,-latex] (2, -2.5) -- (z.south);
    \draw[thick,latex-latex,black!30] (xa) to [out=240,in=280,loop,distance=15pt] (xa);
    \draw[thick,latex-latex,black!30] (xb) to [out=240,in=280,loop,distance=15pt] (xb);
    \draw[thick,latex-latex,black!30] (xc) to [out=240,in=280,loop,distance=15pt] (xc);
    \draw[thick,latex-latex,black!30] (xd) to [out=240,in=280,loop,distance=15pt] (xd);
  \end{tikzpicture}
  \caption{Path diagrams of data generating models. Latent and manifest variables are respectively represented by circles and rectangles. Regression paths and covariances are respectively depicted as single- and double-sided arrows. Nuisance parameters, which are estimated in the first stage of factor score regression, are shown in gray, while focal parameters, which are estimated in the second stage, are shown in black. Panel A: Simple latent regression with continuous manifest variables. Panel B: Latent moderation analysis with continuous manifest variables. Panel C: Multiple latent regression with dichotomous manifest variables.}
  \label{fig:path}
\end{figure}

\subsection{Study 1: Simple Latent Regression with Continuous Manifest Variables}
\subsubsection{Data Generation}
In the first study, the true LV model features a simple latent regression: the predictor and outcome LVs are each measured by five MVs via a linear-normal one-factor model. A path diagram of this model is presented in Figure \ref{fig:path}A. We next describe a DGA for the model that was used throughout Study 1 for data generation from the true population model, bias correction (Algorithm \ref{alg:bc}), and ACM approximation (Algorithm \ref{alg:acm}). For each individual $i$, $i = 1, \dots, n$, let $U_{i1}, \dots, U_{i,12}$ be i.i.d. $\sN(0, 1)$ random variables. We generate the predictor LV $\eta_{i1}$ and the outcome LV $\eta_{i2}$ by
\begin{equation}
  \eta_{i1} = \sqrt{\phi} U_{i1},\ \eta_{i2} = \beta \eta_{i1} + \sqrt{\psi} U_{i2},
  \label{eq:dga1s}
\end{equation}
and subsequently the MVs by
\begin{equation}
  Y_{ij} =  
  \left\{\begin{array}{ll}
      \lambda_j\eta_{i1} + \sigma_jU_{i, j + 2},& j = 1, \dots, 5;\cr
      \lambda_j\eta_{i2} + \sigma_jU_{i, j + 2},& j = 6, \dots, 10.
\end{array}\right.
  \label{eq:dga1m}
\end{equation}
$\lambda_j$ and $\sigma_j^2$ in Equation \ref{eq:dga1m} are the factor loading and unique variance parameters for the $j$th MV, $j = 1,\dots, 10$; $\phi$ in Equation \ref{eq:dga1s} denotes the variance of the predictor LV. These 21 parameters are enclosed in the nuisance parameter vector $\nnu$.\footnote{To identify the outcome measurement model, the variance of the outcome LV must be estimated alongside the factor loadings and unique variances in the first stage. However, this variance is not used for bias correction or subsequent inference. We apply the same treatment in Studies 2 without further comment.} The focal parameter vector $\pphi$ is composed of the remaining two structural parameters in Equation \ref{eq:dga1s}: the latent regression error variance $\psi$ and the regression coefficient $\beta$.

Following \citet{MacCallumEtAl1999}, we varied the true communalities of the five MVs measuring each LV from .7 (high) down to .3 (low) at an interval of .1. Both LVs were standardized: that is, $\phi^* = 1$ and $\beta^*{}^2 + \psi^* = 1$. All the true factor loadings were set to one: $\lambda_j^* = 1$ for all $j = 1,\dots,10$. The unique variances were then determined to attain the prescribed communalities: $\sigma_j^2{}^* = (1 - \rho_j^*{}^2) / \rho_j^*{}^2$, where $\rho_j^*{}^2$ denotes the true communality of the $j$th MV. The true latent regression slope $\beta^*$ was fixed at .6, implying that the predictor LV accounts for 36\% of the variance in the outcome LV. Three sample size conditions were examined: $n$ = 100, 200, and 500.

\subsubsection{Candidate Estimators}

In the first stage of FSR, we estimated separate one-factor models for the two LVs using the R package {\tt lavaan} version 0.6-19 \citep{lavaan}; all models were fitted using ML under the default setting of the {\tt cfa} function. Three types of factor scores were then computed from each unidimensional fitting: the mean score (M), the Bartlett factor score (B), and the regression factor score (R). Under the true LV model, the mean score reliability is .81, following short of the .86 reliability achieved by both the Bartlett and regression factor scores. In the second stage, we regressed the outcome factor scores on the predictor factor scores. Specifically, we always chose the same type of factor scores for both LVs. Because the measurement model is identical for both LVs, using the same scoring method ensures that the estimated regression slopes are comparable to their true values. The resulting FSR solutions are abbreviated as MM (using mean scores), BB (using Bartlett scores), and RR (using regression scores) in the summary of results. To perform bias correction via Algorithm \ref{alg:bc}, we set the number of Robbins-Monro iterations $K = 1000$ and the learning rate constants $a = 3$ and $b = .6$. In addition, a lower bound $10^{-6}$ was imposed for the variance component $\psi$ in Robbin-Monro iterations (see Equation \ref{eq:rm} in Remark \ref{rmk:proj}). For ACM approximation (Algorithm \ref{alg:acm}), we used $M = 1000$ replications and a perturbation constant $\delta = 10^{-6}$.

Two benchmark estimators were included for comparison. First, we obtained the one-stage ML estimator of the full LV model (Figure \ref{fig:path}A) using the {\tt cfa} function in the {\tt lavaan} package, with SEs derived from the expected Fisher information matrix. Second, we applied the bias-avoiding FSR solution proposed by \citet{SkrondalLaake2001}, regressing the Bartlett factor score for the outcome LV on the regression factor score for the predictor LV (henceforth abbreviated as BR). Note that this estimator is consistent only for the latent regression slope but not for the error variance. Hence, we computed parametric bootstrap SEs (based on 1000 Monte Carlo samples) solely for the slope.

\begin{table}[!t]
  \begin{center}
  \caption{Relative bias (RB), empirical standard error (ESE), and relative bias in SE (RBSE) for candidate estimators in Study 1.}
  \label{tab:study1}
  \begin{tabular}{llrrrrrr}
  \toprule
  \multirow{2}{*}{$n$} & \multirow{2}{*}{Method} &
  \multicolumn{3}{c}{$\beta^* = .6$} & 
  \multicolumn{3}{c}{$\psi^* = .64$}\\
  \cmidrule(r){3-5}
  \cmidrule(r){6-8}
     & & RB & ESE & RBSE & RB & ESE & RBSE\\
  \midrule
100 & FSR(MM) &\bf$-$.19 & .09 &      &\bf   .48 & .13 &     \\
    & FSR(BB) &\bf$-$.14 & .10 &      &\bf   .36 & .16 &     \\
    & FSR(RR) &\bf$-$.15 & .08 &      &      .01 & .09 &     \\
    &  BC(MM) & .02 & .12 &  .08 &  .02 & .15 &  .09\\
    &  BC(BB) & .00 & .12 &  .09 &  .04 & .15 &  .06\\
    &  BC(RR) & .00 & .12 &\bf  .11 &  .04 & .15 &\bf  .11\\
    &  ML     & .02 & .12 &  .01 & $-$.01 & .14 &  .02\\
    & FSR(BR) &$-$.01 & .11 &\bf  .13 &\bf  .36 & .16 &\\
  \midrule
200 & FSR(MM) &\bf$-$.20 & .06 &      &\bf  .47 & .10 &     \\
    & FSR(BB) &\bf$-$.15 & .07 &      &\bf  .34 & .12 &     \\
    & FSR(RR) &\bf$-$.15 & .06 &      & $-$.01 & .06 &     \\
    &  BC(MM) & .00 & .09 &  .01 &  .00 & .11 &  .03\\
    &  BC(BB) &$-$.01 & .09 &  .01 &  .01 & .11 &  .01\\
    &  BC(RR) &$-$.01 & .09 &  .02 &  .01 & .11 &  .03\\
    &  ML     & .00 & .08 & $-$.03 & $-$.02 & .11 & $-$.03\\
    & FSR(BR) &$-$.01 & .08 &  .07 &\bf  .34 & .12 &\\
  \midrule
500 & FSR(MM) &\bf$-$.19 & .04 &      & \bf .47 & .06 &     \\
    & FSR(BB) &\bf$-$.15 & .04 &      & \bf .34 & .07 &     \\
    & FSR(RR) &\bf$-$.15 & .04 &      & $-$.02 & .04 &     \\
    &  BC(MM) & .00 & .05 &  .02 &  .00 & .07 & $-$.02\\
    &  BC(BB) & .00 & .05 &  .03 &  .00 & .07 & $-$.01\\
    &  BC(RR) & .00 & .05 &  .04 &  .00 & .07 & $-$.01\\
    &  ML     & .00 & .05 & $-$.01 & $-$.01 & .07 & $-$.05\\
    & FSR(BR) & .00 & .05 &  .10 &\bf  .34 & .07 &\\
  \bottomrule
  \end{tabular}
  \end{center}
  \textit{Note. $n$: Sample size. $\beta^*$: True latent regression slope. FSR: Naive factor score regression. BC: Bias-corrected FSR. MM: Using mean scores for both latent variables (LVs). BB: Using Bartlett factor scores for both LVs. RR: Using regression factor scores for both LVs. BR: Using Bartlett factor scores for the outcome LV and regression factor scores for the predictor LV. ML: Maximum likelihood. RB and RBSE greater than .1 in absolute values are highlighted in bold.}
\end{table}

\subsubsection{Results}
Table \ref{tab:study1} summarizes the results of our first simulation study. As expected, the naive FSR estimators (i.e., MM, BB, and RR) exhibited noticeable bias across most conditions. Regarding the regression slope ($\beta^* = .6$), the naive FSR estimates are attenuated; the MM estimator is more attenuated then the other two, attributable to the lower reliability of the mean score. Because the variances of the factor scores differ from that of the LV itself, the MM and BB estimates for the regression error variance deviate substantially from the true value ($\psi^* = .64$). Meanwhile, the RR estimates of $\psi^*$ are closer, as the regression factor scores are less variable. Regardless of the choices of factor score types, the proposed bias-correction technique successfully reduces the relative bias to less than 5\% across all the conditions, performing comparably to the gold-standard, one-stage ML estimator. While the bias-avoiding BR method accurately recovers the latent regression slope, it yields positively biased error variance estimates similar to the BB method. This later observation is expected as the Bartlett factor scores tend to be more variable than the LV. The biased error variance estimate in the BR solution is a notable limitation. If users compute effect size measures, such as standardized regression slopes and the coefficient of determination, based on the biased variance estimate, their conclusions may be misleading.

Across all sample size conditions, the three bias-corrected estimators and the one-stage ML estimator exhibit equivalent variability for both parameters; for the latent regression slope, their empirical SEs are further comparable to those of the BR estimator. This property appears robust to the choice of factor scores; even with the less reliable mean scores, the bias-corrected MM estimator still achieves efficiency comparable to the ML benchmark. Regarding variance estimation, when $n\ge 200$, the Monte Carlo SEs for the bias-corrected estimator (produced by Algorithm \ref{alg:acm}) closely match the empirical SEs with under 10\% relative bias. For $n = 100$, the estimated SEs for the RR estimator tend to be slightly inflated (approximately 11\% relative bias). We consider this slight SE inflation acceptable, as it results in conservative inference rather than the more problematic liberal inference caused by deflated SEs.

\subsection{Study 2: Latent Moderation Analysis with Continuous Manifest Variables}

\subsubsection{Data Generation}

Next, we study latent moderation model, in which the outcome LV is predicted by the predictor and moderator LVs via a bilinear interaction model. Each LV is measured by five MVs. A path diagram for this model is presented in Figure \ref{fig:path}B. The DGA used in this study utilizes i.i.d. $\sN(0, 1)$ random components $U_{i1}, \dots, U_{i, 18}$ for each individual $i$, $i = 1,\dots, n$. The three LVs are then computed by
\begin{equation}
  \begin{bmatrix}
    \eta_{i1}\\\eta_{i2}
  \end{bmatrix} =  
  \begin{bmatrix}
    \sqrt{\phi_{11}} & 0\\
    \displaystyle\frac{\phi_{21}}{\sqrt{\phi_{11}}} & 
    \displaystyle\sqrt{\frac{\phi_{22}\phi_{11} - \phi_{21}^2}{\phi_{11}}}
  \end{bmatrix}
  \begin{bmatrix}
    U_{i1}\\U_{i2}\\
  \end{bmatrix},\
  \eta_{i3} = \beta_1\eta_{i1} + \beta_2\eta_{i2} + \beta_3\eta_{i1}\eta_{i2} + \sqrt{\psi} U_{i3},
  \label{eq:dga2s}
\end{equation}
in which the $2\times 2$ matrix on the right-hand side of the first equation is the lower-triangular Cholesky factor of the covariance matrix with entries $\phi_{11} = \var(\eta_{i1})$, $\phi_{22} = \var(\eta_{i2})$, and $\phi_{21} = \cov(\eta_{i2}, \eta_{i1})$. Next, the MVs are generated by
\begin{equation}
  Y_{ij} =  
  \left\{\begin{array}{ll}
      \lambda_j\eta_{i1} + \sigma_jU_{i, j + 3},& j = 1, \dots, 5;\cr
      \lambda_j\eta_{i2} + \sigma_jU_{i, j + 3},& j = 6, \dots, 10;\cr
      \lambda_j\eta_{i3} + \sigma_jU_{i, j + 3},& j = 11, \dots, 15.
\end{array}\right.
  \label{eq:dga2m}
\end{equation}
For two-stage estimation, the nuisance parameter vector $\nnu$ is formed by collecting the factor loadings and unique variance parameters for all the 15 MVs, together with the variances of the predictor and moderator LVs $\phi_{11}$ and $\phi_{22}$. The remaining five parameters, including the inter-factor covariance $\phi_{21}$, latent regression coefficients $\beta_1$, $\beta_2$, and $\beta_3$, and the regression error variance $\psi$, compose the focal parameter vector $\pphi$.

We followed the simulation setup in \citet{NgChan2020} to specify the true parameter values for the latent moderation model. The measurement model parameters were selected to yield a coefficient omega (i.e., reliability of the summed or mean score) of .8. To achieve this, we set the true factor loadings to 1, .8, .8, .8, and .8 for the five MVs measuring each LV. The corresponding unique variance parameters took values .44, .66, .88, 1.1, and 1.32. Both the predictor and moderator LVs were standardized (i.e., $\phi^*_{11} = \phi^*_{22} = 1$), and their correlation was set to $\phi^*_{21} = .3$. The true latent regression coefficients were fixed at $\beta^*_1 = \beta^*_2 = .4$ and $\beta^*_3 = .2$. Finally, the error variance $\psi^* = .54$, ensuring a unit variance for the outcome LV. We simulated data with three sample sizes: $n = 100$, 200, and 500.

\subsubsection{Candidate Estimators}

To perform FSR, we followed a procedure similar to that described in Study 1. We estimated separate one-factor models for the three LVs by ML, computed Bartlett factor scores, and obtained OLS estimates for the regression coefficients and error variance. The correlation between the predictor and moderator LVs (i.e., $\phi_{21}$) was estimated by the correlation between their respective factor scores. It is evident from Equation \ref{eq:dga2s} that the outcome LV is no longer normally distributed because its generation involves the product of predictor and moderator LVs. However, ML estimation for the outcome-side measurement model remains $\sqrt{n}$-consistent as the model is identifiable from the covariance structure \citep{BrowneShapiro1988}. Algorithms \ref{alg:bc} and \ref{alg:acm} for point and variance estimation were configured as in Study 1. In addition, we applied projection operators via Equation \ref{eq:rm} to ensure that $|\phi_{21}|\le 1-10^{-6}$ and $\psi\ge 10^{-6}$.

  The latent moderation model is no longer a linear-normal SEM: its (marginal) likelihood function is an intractable integral due to the non-normality of the outcome LV. Numerical integration is typically required to evaluate the likelihood function and perform ML estimation. In this simulation, we used M\textit{plus} version 8.11 \citep{mplus} with the default numerical quadrature and convergence settings to obtain the ML estimator. SEs were computed using the observed Fisher information matrix evaluated at the ML solution.

\begin{table}[!t]
  \begin{center}
  \caption{Relative bias (RB), empirical standard error (ESE), and relative bias in SE (RBSE) for candidate estimators in Study 2.}
  \label{tab:study2}
  \small
  \hspace*{-0.3in}
  \begin{tabular}{llrrrrrrrrrrrr}
  \toprule
  \multirow{2}{*}{$n$} & \multirow{2}{*}{Method} &
  \multicolumn{3}{c}{$\beta_1^* = .4$} & 
  \multicolumn{3}{c}{$\beta_3^* = .2$} &
  \multicolumn{3}{c}{$\psi^* = .54$} &
  \multicolumn{3}{c}{$\phi_{21}^* = .3$} \\
  \cmidrule(r){3-5}
  \cmidrule(r){6-8}
  \cmidrule(r){9-11}
  \cmidrule(r){12-14}
     & & RB & ESE & RBSE & RB & ESE & RBSE & 
     RB & ESE & RBSE & RB & ESE & RBSE\\
  \midrule
100 & FSR     &\bf $-$.14 & .09 &      & \bf $-$.37 & .08 &     &\bf .50 & .16 &      & $-$.01 &  .13 &      \\
    &  BC     &$-$.02 & .12 &  .10 & $-$.06 & .12 & .09 & .05 & .15 &  .00 & $-$.01 &  .13 &   .01\\
    &  ML     &.01 & .12 &  $-$.04 & .02 & .11 & $-$.01 & $-$.05 & .14 & $-$.04 & .02 &  .12 &  .01 \\
    \midrule
200 & FSR     &\bf $-$.13 & .07 &      &\bf $-$.32 & .05 &      & \bf .50 & .11 &      & $-$.01 &  .10 &     \\
    &  BC     &  .00 & .08 &  .05 &  .01 & .08 &  .04 &  .02 & .10 &  .01 & $-$.01 &  .10 & $-$.04\\
    &  ML     &  .01 & .08 & $-$.02 &  .02 & .08 & $-$.03 & $-$.02 & .10 & $-$.04 & $-$.01 &  .10 &  $-$.08\\
    \midrule
500 & FSR     &\bf $-$.13 & .04 &      &\bf $-$.33 & .03 &      & \bf .49 & .07 &      & $-$.01 &  .06 &     \\
    &  BC     &  .00 & .05 &  .01 & $-$.01 & .05 &  .07 &  .00 & .06 &  .02 & $-$.01 &  .06 & $-$.03\\
    &  ML     &  .00 & .05 & $-$.03 &  .00 & .05 &  .00 & $-$.02 & .06 & $-$.01 & $-$.01 &  .06 &  $-$.06\\
  \bottomrule
  \end{tabular}
  \end{center}
    \textit{Note. $n$: Sample size. $\beta_1^*$: True partial effect for predictor latent variable (LV). $\beta_3^*$: True interaction effect. $\psi^*$: True error variance for latent regression. $\phi_{21}^*$: True correlation between the predictor and moderator LVs. FSR: Naive factor score regression using Bartlett factor scores. BC: Bias-corrected FSR. ML: Maximum likelihood. RB and RBSE greater than .1 in absolute values are highlighted in bold.}
  \end{table}

  \subsubsection{Results}

  Results for Study 2 are shown in Table \ref{tab:study2}. Given the symmetry in measurement models and true coefficients for the predictor and moderator LVs, we omit results for the partial effect $\beta_2$. We again observe that the FSR estimates for the partial and interaction effects are substantially biased, with relative bias reaching 13\% and 33\%, respectively. The error variance $\psi$ is also poorly recovered by FSR (approximately 50\% relative bias); as in Study 1, this is likely attributed to the fact that Bartlett factor scores are more variable than LVs themselves. Meanwhile, FSR accurately estimates the between-LV correlation parameter $\phi_{21}$ (approximately 1\% relative bias) in all three sample sizes. In contrast, the proposed bias-correction algorithm restrains relative bias to under 5\% for all focal parameters, a performance on par with the one-stage ML benchmark.

  We also note that the statistical efficiency of the bias-corrected FSR estimator is nearly optimal, evidenced by empirical SE values indistinguishable from the one-stage ML estimator. Moreover, Algorithm \ref{alg:acm} yields accurate SE estimates for all biased-corrected focal parameter estimates, keeping the relative bias in SE below 10\%. The largest SE overestimation occurs for the partial effect $\beta_1$ at $n = 100$. As discussed earlier in Study 1, overestimating sampling variability leads to conservative inference; while this may slightly reduce statistical power, the false positive rate remains under control. 

  \subsection{Study 3: Multiple Latent Regression with Dichotomous Manifest Variables}

  \subsubsection{Data Generation}

  In the final study, we examine a multiple latent regression model wherein the four predictor LVs and one outcome LV are jointly measured by an independent-cluster IRT model (see Figure \ref{fig:path}C for its path diagram). The structural part of the model is characterized by the following DGA:
  \begin{equation}
    \begin{aligned}
      \begin{bmatrix}
        \eta_{i1}\\ 
        \eta_{i2}\\ 
        \eta_{i3}\\ 
        \eta_{i4}
      \end{bmatrix} &= {\underbrace{\begin{bmatrix}
        1\\
        \phi_{21} & 1\\
        \phi_{31} & \phi_{32} & 1\\
        \phi_{41} & \phi_{42} & \phi_{43} & 1\\
  \end{bmatrix}}_{\PPhi}}^{1/2}
      \begin{bmatrix}
        U_{i1}\\
        U_{i2}\\
        U_{i3}\\
        U_{i4}
      \end{bmatrix},\
      \eta_{i5} &=  \beta_1\eta_{i1} + \beta_2\eta_{i2} + \beta_3\eta_{i3} + \beta_4\eta_{i4} + \sqrt{\psi}U_{i5},
    \end{aligned}
    \label{eq:dga3s}
  \end{equation}
  in which $U_{i1}, \dots, U_{i5}$ are i.i.d. $\sN(0, 1)$ variates, and $\PPhi^{1/2}$ denotes a $4\times 4$ matrix satisfying $\PPhi = \PPhi^{1/2}(\PPhi^{1/2})\t$ (e.g., the lower-triangular Cholesky factor of $\PPhi$). Each LV is indicated by eight unique two-parameter logistic (2PL) items \citep{Birnbaum1968} via the Bernoulli DGA:
  \begin{equation}
    Y_{ij} = 1\left\{\mathrm{logit}(U_{i,\lceil j/8\rceil})\le \alpha_j + \gamma_j\eta_{i,\lceil j/8\rceil}\right\},
    \label{eq:dga3m}
  \end{equation}
  in which $U_{i,j+5}$, $j = 1, \dots, 40$, are i.i.d. $\sU(0, 1)$ variates, $\mathrm{logit}(u) = \log(u/(1 - u))$, $u\in(0, 1)$, denotes the logit link function, and $1\{A\}$ is the indicator function for set $A$. Parameters in Equations \ref{eq:dga3s} and \ref{eq:dga3m} are partitioned into the nuisance and focal subsets for two-stage FSR: the nuisance parameter vector $\nnu$ comprises item intercepts ($\alpha_j$'s) and slopes ($\gamma_j$'s), and the focal parameter vector $\pphi$ is composed of the four latent regression slopes ($\beta_1, \dots, \beta_4$) and the six correlation parameters among the predictor LVs ($\phi_{21}, \phi_{31}, \dots, \phi_{43}$).\footnote{Unlike studies 1 and 2, the LV variance is fixed to 1 in the estimation of each 2PL model; therefore, no variance component is estimated as a nuisance parameter and the latent regression error variance $\psi$ is uniquely determined given the regression slopes.}

  The following true parameter values were used in the simulation. For each LV, the eight item slopes were set to 1, 1.25, \dots, 1.75. Item intercepts were computed by $\gamma_j^* = -\alpha^*_j\delta_j^*$ for each $j$, where $\delta_j^*$ indicates the true difficulty parameter for the $j$th item. The true item difficulty values ranged from $-1.75$ to 1.75 at an interval of .5. In the structural model, the true latent regression coefficients were set to $\bbeta^* = (\beta_1^*, \beta_2^*, \beta_3^*, \beta_4^*)\t = (.1, .2, .3, .4)\t$, and the true correlations among the predictor LVs $\phi_{21}^*=\phi_{31}^*=\phi_{43}^* = .3$. To ensure standardized outcome LVs, we set the error variance  to $\psi^* = 1 - \bbeta^*{}\t\PPhi^*\bbeta^*$ = .49. Given the complexity of the model, we increase the sample sizes in Study 3 to $n = 500$ and 1000.

  \subsubsection{Candidate Estimators}
  Unidimensional 2PL models were estimated for one LV at the time during the first stage of FSR. ML estimates of the item intercepts and slopes (i.e., nuisance parameters) were obtained using the R package \texttt{mirt} version 1.39 with the default convergence and numerical quadrature  settings \citep{mirt}. Expected \emph{a posteriori} factor scores were then computed for each LV, and an OLS regression was subsequently fitted to the factor scores to obtain estimates of latent regression slopes. Between-LV correlations were estimated by the corresponding sample correlations among predicted factor scores. The configurations of Algorithms \ref{alg:bc} an \ref{alg:acm} were more involved in this study, which we elaborate next. 

  Unlike Studies 1 and 2, where only one or two parameters were bounded, here the predictor LV correlation matrix $\PPhi$ must remain positive semi-definite, while the latent regression slopes $\bbeta$ must satisfy the constraint $1 - \bbeta\t\PPhi\bbeta\ge 0$. Orthogonal projection onto the joint feasible sets for $\PPhi$ and $\bbeta$ proves to be challenging even numerically. To prevent invalid iterates, we apply an alternative two-step operation after updating $\PPhi$ and $\bbeta$: (1) truncating eigenvalues of $\PPhi$ at the lower bound $10^{-6}$ and standardizing the resulting positive definite matrix to a correlation matrix $\tilde\PPhi$, and (2) solving the quadratic programming problem $\min_{\tilde\bbeta}(\tilde\bbeta - \bbeta)\t(\tilde\bbeta - \bbeta)$ subject to $1 - \tilde\bbeta\t\tilde\PPhi\tilde\bbeta\ge 10^{-6}$. Pilot simulations indicated that this intervention is only occasionally performed in the first few iterations: as the learning rate decays, the iterates tend to stay within bounds naturally. The learning rate constants were set to $a = 3$ and $b = .6$, the same as in Studies 1 and 2.

  For Algorithm \ref{alg:acm}, we adopted the blockwise SP strategy described in Remark \ref{rmk:blk}. Pilot simulations suggest that more than 5000 Monte Carlo samples would have been needed to accurately approximate the SEs had we applied SP to all 90 model parameters simultaneously. Specifically, we partitioned the parameters into 15 equal-sized blocks (so that each block contains six parameters) and executed SP within each block to obtain Jacobian submatrices. Regarding the other tuning aspects, the number of Monte Carlo replications was set to $M = 1000$ (consistent with Studies 1 and 2), and the perturbation constant was fixed at $\delta = .005$.

  Two $\sqrt{n}$-consistent estimators were included as benchmarks for comparison. The one-stage ML estimator was obtained using an implementation of the Metropolis-Hastings Robbins-Monro (MH-RM) algorithm from the R package \texttt{mirt}. MH-RM is a generic SA algorithm developed to handle intractable marginal likelihood for complex LV models \citep{Cai2010a, Cai2010b}. While most default settings of the MH-RM algorithm were preserved, we modified the Metropolis-Hastings (MH) proposal standard deviation to .3; we also extended the number of burn-in iterations to 500 and the number of random-walk Metropolis draws to 1000. Seeing that the \texttt{mirt} package cannot handle regression among LVs, we first fitted the equivalent five-dimensional independent-cluster IRT model and then transformed the between-LV correlations to regression coefficients. The corresponding SEs for focal parameters were obtained by the Delta method. In addition, we obtained the diagonally weighted least square (DWLS) estimator using M{\it plus} (i.e., \texttt{ESTIMATOR = WLSMV} and \texttt{PARAMETERIZATION = THETA}) with default settings. DWLS is a widely used multi-stage, limited-information estimation method for IRT models, in which the focal parameters are estimated conditional on sample estimates of item thresholds and tetrachoric correlations.\footnote{Strictly speaking, DWLS estimates a slightly different IRT model, replacing the logit link in Equation \ref{eq:dga3m} by a probit link. However, it is generally accepted that the two link functions resemble each other closely, differing only in the scale of item parameters.}

  \begin{sidewaystable}
    \begin{center}
    \caption{Relative bias (RB), empirical standard error (ESE), and relative bias in SE (RBSE) for candidate estimators in Study 3.}
    \label{tab:study3}
    \small
    \begin{tabular}{llrrrrrrrrrrrrrrr}
    \toprule
    \multirow{2}{*}{$n$} & \multirow{2}{*}{Method} &
    \multicolumn{3}{c}{$\beta_1^* = .1$} & 
    \multicolumn{3}{c}{$\beta_2^* = .2$} &
    \multicolumn{3}{c}{$\beta_3^* = .3$} &
    \multicolumn{3}{c}{$\beta_4^* = .4$} &
    \multicolumn{3}{c}{$\phi_{21}^* = .3$} \\
    \cmidrule(r){3-5}
    \cmidrule(r){6-8}
    \cmidrule(r){9-11}
    \cmidrule(r){12-14}
    \cmidrule(r){15-17}
       & & RB & ESE & RBSE & RB & ESE & RBSE & RB & ESE & RBSE & 
       RB & ESE & RBSE & RB & ESE & RBSE\\
    \midrule
  500 & FSR & .08 &.04 &    &\bf $-$.16 &.04 &    &\bf $-$.24 &.04 &    &\bf$-$.27  &.04 &     &\bf$-$.32 & .04 &    \\
      &  BC &$-$.03 &.07 &.02 &$-$.02 &.07 &.04 &$-$.02 &.06 &.07 & .01  &.06 & .06 &$-$.01 & .06 & .01\\
      &  ML &$-$.04 &.07 &.06 &$-$.04 &.07 &\bf.11 & .01 &.07 &.08 & .05  &.07 & .07 & .03 & .07 & .04\\
      &DWLS &$-$.05 &.07 &$-$.06&$-$.02 &.07 &$-$.05&$-$.01&.07 &$-$.02 &.02&.06&$-$.01 &.03 & .07 &$-$.06\\
      \midrule
  1000& FSR &\bf .10 &.03 &    &\bf$-$.16 &.03 &    &\bf$-$.23 &.03 &    &\bf$-$.28  &.03 &     &\bf$-$.33 & .03 &    \\
      &  BC & .00 &.05 &.05 &$-$.01 &.05 &.02 & .00 &.04 &.07 &$-$.01  &.05 &$-$.05 &$-$.02 & .04 &$-$.01 \\
      &  ML &$-$.04 &.05 &\bf.18 & .00 &.05 &\bf.15 & .02 &.05 &\bf.17 & .01  &.05 & .10 & .00 & .05 & .08 \\
      &DWLS &$-$.01 &.05 &.00 & .00&.05&$-$.02 & .01&.04&.01 &.00&.05&$-$.09 & $.01$ &.05 &$-$.04\\
    \bottomrule
    \end{tabular}
    \end{center}
    \textit{Note. $n$: Sample size. $\beta_1^*$: True partial effect for predictor latent variable (LV). $\beta_3^*$: True interaction effect. $\psi^*$: True error variance for latent regression. $\phi_{21}^*$: True correlation between the predictor and moderator LVs. FSR: Naive factor score regression using Bartlett factor scores. BC: Bias-corrected FSR. ML: Maximum likelihood. DWLS: Diagonally weighted least square. RB and RBSE greater than .1 in absolute values are highlighted in bold.}
\end{sidewaystable}

\subsubsection{Results}

Table \ref{tab:study3} shows that the naive FSR estimates for latent regression coefficients are biased as expected. Because the LVs are correlated, this bias is not always attenuation (i.e., negative bias since the true effect is positive). For instance, when $\beta_1^* = .1$ at $n = 1000$, the FSR estimates are inflated by approximately 10\%. From $\beta_2^*$ to $\beta_4^*$, the relative bias increases with the magnitude of the true coefficient, approaching 28\% for $\beta_4^* = .4$. Meanwhile, the correlations among factor scores are significantly attenuated, exhibiting more than 30\% downward bias relative to the true between-LV correlation (.3). The proposed bias-correction procedure (Algorithm \ref{alg:bc}) nearly eliminates the bias for all focal parameters across both sample size conditions. Furthermore, both benchmark estimators show negligible bias, consistent with the asymptotic theory.

It is also observed that the bias-corrected FSR, one-stage ML, and multi-stage DWLS estimators show comparable empirical SEs. Theoretically, one-stage ML estimation attains the Cram\'er-Rao lower bound (i.e., has the lowest asymptotic variance) in the limit; however, this does not always translate into a gain in statistical efficiency compared to asymptotically less efficient options in finite samples. The practical utility of one-stage ML is further limited by the intractability of the likelihood and the reliance on SA (e.g., MH-RM) algorithms when the latent dimensionality is high. Additionally, SE estimates for the biased-corrected FSR (via Algorithm \ref{alg:acm}) and DWLS are generally accurate. In contrast, we observe slight inflations in the estimated SEs for the latent regression slopes, which is likely attributed to the application of the Delta method.

\section{Discussion}

In the present paper, we developed a generic bias-correction procedure for two-stage estimation of parametric statistical models and tailored it to FSR: a method widely utilized in psychological research to study relations among unobservable constructs operationalized as LVs. Theoretically, we established the $\sqrt{n}$-consistency of the bias-corrected FSR estimator under mild regularity conditions. On the computational side, we proposed an SA-based point estimation algorithm (Algorithm \ref{alg:bc}) and a Monte Carlo-based algorithm for variance estimation (Algorithm \ref{alg:acm}), both of which are devised to minimize reliance on analytical derivations and thus can be broadly applied. Finally, we demonstrated through a sequence of three Monte Carlo experiments that the bias-corrected FSR estimator performs comparably to the gold-standard one-stage ML estimator in terms of both accuracy and statistical efficiency.

Our method and empirical findings carry two important implications. First, we advocate for viewing FSR as a practical and effective methodological tool, rather than merely criticizing its inherent bias; the latter is a common perspective in the extant methodological literature. We demonstrate that although the bias in naive FSR estimates can be substantial for certain coefficients, it can be removed with ease using the proposed SA-based algorithm. In fact, the resulting bias-corrected FSR estimator recovers focal model parameters with accuracy and efficiency comparable to the asymptotically optimal one-stage ML estimator. Moreover, the empirical performance of the bias-corrected estimator appears agnostic to the specific choice of factor scores (as evidenced in Simulation Study 1). This property is desirable from a practical standpoint, as it relieves substantive researchers of the burden to ``choose the right types of scores'' and thereby simplifies the modeling workflow in validity studies.

The second implication calls for a re-evaluation on the role of bias in assessing point estimators. Conventionally, an estimator is considered acceptable only if it exhibits negligible bias. However, our bias-correction result (Proposition \ref{prop:bc}) indicates that this strict requirement can be considerably relaxed. So long as an estimator responds to local perturbations near the true data-generating parameters, it is eligible for modification into an asymptotically unbiased (or more precisely, a $\sqrt{n}$-consistent) version. This realization suggests an effective strategy for constructing  valid estimators: modify naive estimators that are easy to compute but inaccurate, ultimately achieving an optimal balance between accuracy and computational efficiency.

Several extensions and limitations remain to be addressed in future work. First, the bias-correction and ACM approximation algorithms are broadly applicable to problems where one-stage estimation is numerically challenging, specifically when the likelihood surface is ill-conditioned or multimodal. For instance, \citet{LuoEtAl2025} applied a special case of the proposed method (without nuisance parameters; see Remark \ref{rmk:nonuis}) to the generalized linear local approximation (GLLA) estimator for differential equation models. They demonstrated that the bias-corrected GLLA estimator is both accurate and efficient, outperforming the naive GLLA and one-stage ML in various data-generating conditions. This application highlights the method's general utility in refining estimators that are non-iterative, stable, yet biased. We encourage future studies to evaluate the proposed method in other model families facing similar estimation challenges.

Second, although Algorithms 1 and 2 proved effective in all three simulation studies of the current work, it remains uncertain whether they can scale to more complex LV models (e.g., models with many LVs and nonlinear structural regressions) or maintain satisfactory performance in more demanding estimation scenarios (e.g., small sample sizes). Incorporating second-order information via stochastic quasi-Newton methods \citep[e.g.,][]{WangEtAl2017} could further enhance computational efficiency. Consequently, large-scale numerical experiments are necessary to comprehensively evaluate the proposed methods across a broad range of model specifications and estimation settings.

Third, a possible avenue of future work is to extend the proposed algorithm to handle bias correction in nonparametric and high-dimensional inference. In nonparametric or semiparametric models, the parameter space is infinite-dimensional; parametric approximations are misspecified and thus lead to biased functional estimates. It is of interest to investigate whether modern SA algorithms in functional spaces, such as the kernel or sieve SA \citep{DieuleveutBach2016, ZhangSimon2022}, can be adapted to correct the misspecification bias. Another relevant scenario concerns high-dimensional problems where both the number of parameters and the sample size tend to infinity. While regularization methods (e.g., with ridge or LASSO penalties) are standard in such settings, the induced shrinkage bias must be removed in order to produce valid inference on low-dimensional focal parameters \citep[e.g.,][]{ChernozhukovEtAl2018, NingLiu2017,ZhangZhang2014}. The proposed algorithm may offer a simple computational alternative to existing de-biasing strategies for regularized estimators.

Fourth, ongoing work aims to develop a user-friendly software package implementing the proposed algorithms. For any initial two-stage estimator conforming to the general framework in Section ``Statistical Framework of Two-Stage Estimation'', the corresponding bias correction and variance estimation algorithm require only three user-specified functions: a DGA, a $\sqrt{n}$-consistent nuisance parameter estimator, and an initial focal parameter estimator that is $\sqrt{n}$-consistent for its own expectation. This simple, modular structure facilitates the development of generic code to accommodate a wide variety of models and estimators. While allowing users to customize the component functions is essential, providing example implementations for standard models, such as the three FSR examples discussed in the current work, would significantly enhance the package's accessibility and pedagogical value.

In sum, the present work demonstrates that the intuitive ``measure-then-structural'' estimation strategy, long preferred by applied researchers for its conceptual simplicity, need not come at the cost of estimation accuracy. The proposed bias-correction algorithm effectively removes the asymptotic bias of FSR, transforming this convenient heuristic into a valid estimator for LV models. The resulting bias-corrected estimator thus enjoys the distinct advantages of both paradigms: the implementation convenience of two-stage methods and the theoretical rigor of one-stage methods. It potentially empowers researchers to examine a wider array of complex relationships among latent constructs.

\bibliographystyle{apacite}
\bibliography{TwoStage}
\end{document}